\documentclass[conference]{IEEEtran}
\IEEEoverridecommandlockouts
\usepackage{cite}
\usepackage{amsmath,amssymb,amsfonts}
\usepackage{algorithmic}
\usepackage{graphicx}
\usepackage{textcomp}
\usepackage{xcolor}
\def\BibTeX{{\rm B\kern-.05em{\sc i\kern-.025em b}\kern-.08em
    T\kern-.1667em\lower.7ex\hbox{E}\kern-.125emX}}
\begin{document}

\title{Provisioning Fog Services to 3GPP Subscribers: Authentication and Application Mobility\\

}

\author{\IEEEauthorblockN{Asad Ali\textsuperscript{1},
        Tushin Mallick\textsuperscript{2},
        Sadman Sakib\textsuperscript{2},
        Md. Shohrab Hossain\textsuperscript{2},
        and~Ying-Dar Lin\textsuperscript{1}}
\IEEEauthorblockA{\textit{\textsuperscript{1} Department of Computer Science, National Yang Ming Chiao Tung University, Taiwan} \\
\textit{\textsuperscript{2} Department of Computer Science and Engineering, Bangladesh University of Engineering and Technology, Bangladesh}\\
\small ali.eed06g@nctu.edu.tw, \{tushin.mallick2010, saadsakib3\}@gmail.com, mshohrabhossain@cse.buet.ac.bd, ydlin@cs.nctu.edu.tw}
}

\maketitle

\begin{abstract}
Multi-Access Edge computing (MEC) and Fog computing provide services to subscribers at low latency. There is a need to form a federation among 3GPP MEC and fog to provide better coverage to 3GPP subscribers. This federation gives rise to two issues\textemdash third-party authentication and application mobility\textemdash for continuous service during handover from 3GPP MEC to fog without re-authentication. In this paper, we propose: 1) a proxy-based state transfer and third-party authentication (PS3A) that uses a transparent proxy to transfer the authentication and application state information, and 2) a token-based state transfer and proxy-based third-party authentication (TSP3A) that uses the proxy to transfer the authentication information and tokens to transfer the application state from 3GPP MEC to the fog. The proxy is kept transparent with virtual counterparts, to avoid any changes to the existing 3GPP MEC and fog architectures. We implemented these solutions on a testbed and results show that PS3A and TSP3A provide authentication within 0.345\textendash2.858s for a 0\textendash100 Mbps proxy load. The results further show that TSP3A provides application mobility while taking 40–52\% less time than PS3A using state tokens. TSP3A and PS3A also reduce the service interruption latency by 82.4\% and 84.6\%, compared to the cloud-based service via tokens and prefetching.
\end{abstract}


\section{Introduction}
\noindent\emph{Multi-Access Edge and Fog Computing Federation}

Multi-Access Edge Computing (MEC) and fog computing bring computational capabilities closer to the users to fulfill the requirements of time-sensitive applications. MEC has been standardized by the European Telecommunications Standards Institute (ETSI) to provide computing services to the subscribers of third generation partnership project (3GPP) cellular networks \cite{hu2015mobile} by integrating the computational capabilities of servers with existing 3GPP cellular networks \cite{siriwardhana2021survey} at reduced latency. Fog computing, on the other hand, is another, similar computing paradigm that provides computing and storage functions closer to users \cite{yi2015survey} and adds a fog layer between end devices and cloud layer. The difference between MEC and fog computing is that MEC services are provided by the 3GPP cellular networks while, fog computing services can be deployed by an individual in a smart house or by any company in a smart city environment.

There are different 3GPP Mobile Network Operators (MNOs) around the globe and they have deployed MEC servers in their infrastructure to provide these services to their subscribers. Currently, individual MNOs are not able to provide the MEC coverage to 3GPP subscribers in all areas. Therefore, the MECs deployed by the MNOs are not enough to fulfil the requirements of the 3GPP subscribers. This coverage issue can be solved by provisioning the fog services to the 3GPP subscribers. Fog service providers already exist in smart homes and smart city environments and there is a need of a federation among the MECs and these fog service providers. This 3GPP MEC-Fog Federation is a collaboration between a telecommunication operator and a fog service provider, where they share resources among themselves and provide computing services to their users. Such a federation will be useful for both  service providers and subscribers as service providers will be able to extend their capacity and capabilities while subscribers will enjoy services from multiple providers. \\

\noindent\emph{Authentication and Application Mobility Issues}

The federation we are proposing allows 3GPP subscribers to access the services provided by a fog. However, a few issues arise as a result of such a federation. Subscribers of 3GPP will need to authenticate themselves with the fog service providers in order to access their services. If a user creates a new account to access fog, it will not be feasible as the user would have to buy subscriptions from multiple fog service providers. The solution to this issue is  third-party authentication, where subscribers are able to authenticate themselves with  fog service providers via their 3GPP credentials. The major issue that arises here is that the 3GPP MEC and the fog belong to different trust domains and use different protocols for user authentication, and the message flows of these protocols are different. This gives rise to the \textit{third-party authentication} issue and it becomes necessary to design a solution by which  subscribers (the first party) of 3GPP could authenticate themselves with multiple fog service providers (the second party) using their 3GPP credentials (the third party).

When a user moves out of the MEC coverage of a 3GPP network, it can access the services provided by neighbouring fog service providers. If the user does not move instantaneously into the coverage of neighboring fog service providers, a discontinuation of service would occur, which would increase  latency and degrade the user’s experience. Also, whenever a user moves from a 3GPP MEC services to the fog service providers, active application sessions must be retained so that the user does not have to start a new session at the fog service provider. The application state of active application sessions must be kept intact and transferred to the fog service providers with  minimum latency so that user’s experience is not degraded. This leads us to an \textit{application mobility issue} and we need to design a solution that transfers the session state of users from the 3GPP MEC to the fog servers. In summary, we have identified two major issues that need to be resolved in order to realize a federation among the 3GPP MEC and fog service providers: Third-party authentication and application mobility. \\ 

\noindent\emph{Proxy and Token based Solution}

In order to solve the third-party authentication and application mobility issues, we propose two solutions to end users, namely: 1) Proxy-based state transfer and third-party authentication (PS3A), and 2) Token-based state transfer and Proxy-based third-party authentication (TSP3A). The reason behind proposing these two solutions is to test the efficacy of both the token-based and proxy-based approaches to see which solution is useful under what conditions. PS3A and TSP3A both make use of a transparent proxy to transfer  authentication information of  3GPP subscribers to  fog servers. The basic design idea behind the proxy is transparency by using virtual counterparts to avoid any changes to the message flows of authentication protocols and existing MEC and fog servers’ infrastructures. The two solutions differ from each other in terms of state transfer method. In PS3A, application state transfer is carried out through proxy and, in TSP3A, application state is transferred through a state token. We deployed a testbed to evaluate our proposed solutions and ran experiments to analyse the efficacy of these proposals. The essence of this paper are summarized as follows:

\begin{itemize}
\item We propose two solutions that allow a user to access a fog with a 3GPP subscription by using a transparent proxy and token based approach.
\item The proposed proxy provides translation among multiple authentication protocols and transfers a user’s authentication information across different trust domains.
\item The proposed token-based method transfers the application state information across different domains at much reduced latency.
\end{itemize}

\begin{table*}[!t]
\renewcommand{\arraystretch}{1.3}
\caption{Related Work}
\label{table_one}
\centering
\begin{tabular}{|c||c||c||c||c|c|}
\hline
Name & Method & Federation Scenario & 3p/Multi-Network & Application & Transparency\\
 & & F=Fog, E= MEC/Cellular & Authentication & Mobility & \\
\hline

Albarki \cite{albakri2018hierarchical}, & Centralized authentication server & F only & \checkmark & X& X\\
Chen \cite{chen2020secure} & & & & & \\
\hline
3GPP TS 33.402 \cite{3gpptech2020} & EAP-AKA & E-F & \checkmark & X & \checkmark\\
\hline
Hotspot 2.0 \cite{hotspot2015} & EAP-AKA & E-F & \checkmark & X & \checkmark\\
\hline
Shidhani \cite{al2010fast}, & Modified EAP-AKA & E-F & \checkmark & X & X\\ 
Hyeran \cite{mun20093g} & & & & &\\
\hline
Minghui \cite{shi2007service} & Integrated service model & E-F & \checkmark & X & X\\
\hline
Live Service Migration \cite{machen2017live}& Layered framework & E-E &X&\checkmark&X\\
\hline
FAST \cite{doan2021fast}& SDN & E-E & X & \checkmark & \checkmark\\
\hline
Santos \cite{santos2019towards}, & REST HTTP web service & F-F & X & \checkmark & \checkmark\\
FogBus \cite{tuli2019fogbus} & & & & &\\
\hline
Ours & Transparent Proxy/token & E-F & \checkmark & \checkmark & \checkmark\\
\hline
\end{tabular}
\end{table*}

\section{Related Work}
We examined the literature that  propose solutions to either third-party authentication or application mobility issue in  3GPP MEC or fog. There are various studies that address authentication and application mobility and some of these propose novel authentication methods with a central authentication server in fog networks \cite{albakri2018hierarchical} \cite{chen2020secure}. Since, these authentication methods are not standard, these are not suitable for a 3GPP-MEC federation. For the users switching from 3GPP network to a different wireless network, some studies propose EAP-based methods [\cite{3gpptech2020}, \cite{hotspot2015}, \cite{al2010fast}, \cite{mun20093g}]. EAP-based methods are executed from fog nodes and suffer from unnecessary authentication delay in the case of frequent handovers in a fog network. Another study proposes an integrated service model for roaming between cellular network and wireless LAN \cite{shi2007service}.

Regarding application mobility, \cite{machen2017live} proposes  the stateful migration of service applications between two edge clouds while considering an application instance consisting of three layers, of which only absent layers are moved to the target platform. In another study FAST \cite{doan2021fast} forwards states from source a instance to a destination instance using a programmable state forwarding framework based on Software-Defined Networking (SDN). Adopting these strategies requires conforming to  methods in both source and target platforms which might not be possible in a federation between different trust domains. Santos \cite{santos2019towards} and FogBus \cite{tuli2019fogbus} use Representational State Transfer (REST) procedures in HTTP to share information between two platforms to transfer application information. \cite{ali2020transparent} provides a token-based solution to these issues strictly within the MECs deployed by a single 3GPP cellular network (Intra-MNO MEC environment) and another study \cite{lin2020proxy} proposes a transparent proxy that solves the third-party authentication issue in a federated 3GPP cellular edge and cloud environment. To the best of our knowledge, this ours the first study that addresses the two key issues in  service resumption for mobile users switching from a 3GPP MEC network to a fog network. We provide a simple and transparent solution that can be adapted for any federation between a fog network and a cellular provider.

\section{Problem Formulation}
\subsection{Problem Scenarios}

\begin{figure}[!t]
\centering
\includegraphics[width=3.4in]{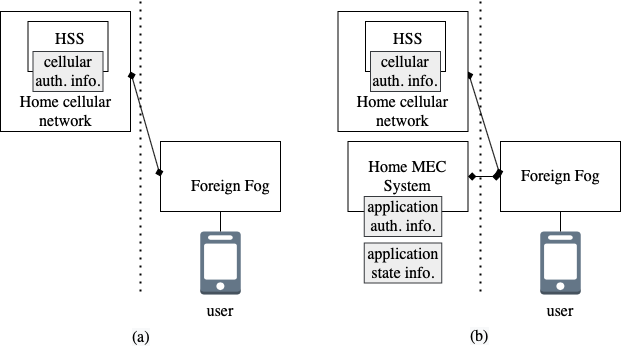}
\caption{(a) Authentication problem for new application (b) Authentication and application mobility problems for application continuation.}
\label{fig_one}
\end{figure}

Consider an MEC deployed in a 3GPP MNO which is in a federation with a fog service provider. We assume that a user is the subscriber of the 3GPP MNO and uses the services provided by the MEC deployed by the MNO. The home subscriber server (HSS) in the 3GPP MNO contains the subscription and authentication information of the user; the fog service provider does not have any information of the user. In this paper, we consider two scenarios: in the first one, we assume that the user wants to access the services provided by the fog service provider and needs to become authenticated with that fog service provider via the authentication credentials from the 3GPP MNO network (Fig. \ref{fig_one}(a)). In the other scenario,  shown in Fig. \ref{fig_one}(b), we assume that the user has moved to the fog service provider while using the MEC services in the 3GPP MNO. In this scenario, the fog service provider needs to get the authentication information from the 3GPP MEC, along with the application state information in order to obtain the service continuity.  

We will only consider the scenario where the UE moves from the 3GPP MNO to the fog service provider, because our objective is to provide fog services to the 3GPP subscribers by performing transparent 3rd-party authentication along with seamless application mobility with minimal latency. An application can be stateful or stateless \cite{machen2017live}. In stateful applications, user data, also called an application state, denotes application usage, such as the number of seconds watching a video. In such applications, when a user switches the service providers, the application state has to be migrated to resume the application from same position while keeping service interruption delay to a minimum.

\subsection{Problem Statement}

We have an MEC framework in a 3GPP cellular network that is federated with a fog network. A 3GPP subscriber is authenticated with the 3GPP cellular network and it may or may not be using certain applications in the 3GPP MEC. The subscriber moves to a fog network and wants to access applications in the fog server. The objective is to provide the services of the fog to the subscriber without creating another account. We assume that the fog is using OpenID Connect (OIDC), which is a popular third-party authentication mechanism that allows a client to authenticate an end-user based on authentication with an authorization server and obtain information about user \cite{sakimura2014openid}. It is predicted that in the coming years, OIDC will have widespread adoption in fog computing and IoT applications \cite{navas2019understanding}. The subscriber must also start using a particular application in the fog network from the same state it had been left off in the 3GPP MEC. This objective must be achieved at low latency while maintaining the transparency of existing 3GPP and fog architecture and protocols.

\section{Proposed Architecture Design}
We propose Proxy-based state transfer and third-party authentication (PS3A), and Token-based state transfer and Proxy-based third-party authentication (TSP3A) for solving the authentication and application mobility problems. PS3A and TSP3A make use of a transparent proxy to transfer the user’s information from the 3GPP MEC to the fog. The major design idea behind the proxy is transparency so as to avoid any modifications in the existing 3GPP cellular network, MEC, and fog infrastructure. We provide transparency by proposing virtual counterparts inside the proxy to communicate  the MEC and fog entities with their virtual counterparts. PS3A and TSP3A share a common third-party authentication solution, via proxy, and differ in the state transfer method for application mobility via the proxy and via the token, respectively.

\begin{figure}[!t]
\centering
\includegraphics[width=3.5in]{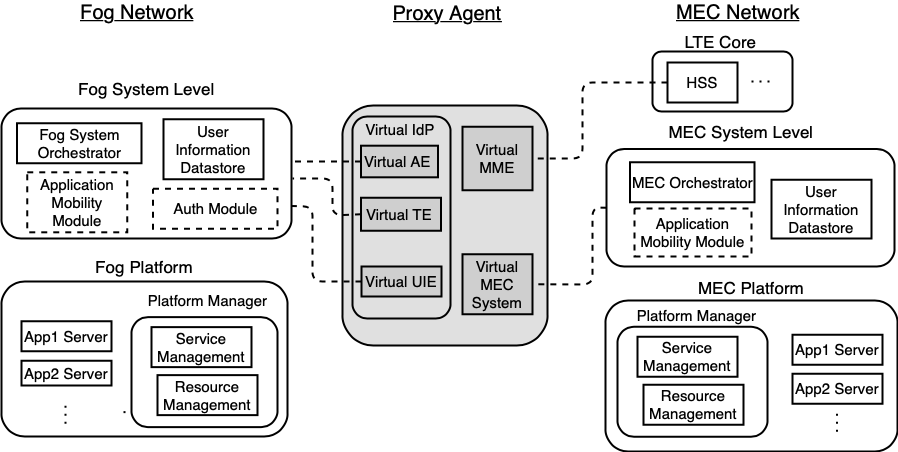}
\caption{PS3A and TSP3A Architecture.}
\label{fig_two}
\end{figure}

\subsection{Architecture}
The proxy needs to be deployed between the 3GPP MEC and fog network, as shown in Fig. \ref{fig_two}. The MEC platform is deployed in a 3GPP cellular network that contains the necessary infrastructure to run MEC applications \cite{etsi2019multi}. The MEC platform manager takes care of the application requirements and a system level entity coordinates all MEC platforms within the 3GPP MEC network via a system orchestrator. In the fog network, an authentication module handles the authentication related tasks. Application Mobility Module in both fog and MEC systems handles the tasks related to the application state transfer. The proposed proxy connects the fog and MEC network at system level using different virtual counterparts. We assume that OIDC is available as an authentication mechanism in the fog network. Therefore, the proxy acts as a virtual Identity Provider (vIdP) while communicating with the fog so that the Relying party (RP) component in the fog can communicate with vIdP. The vIdP consists of a virtual Authorization Endpoint (vAE), a virtual Token Endpoint (vTE) and a virtual Userinfo Endpoint (UIE). The proxy acts as a virtual MME (vMME) and a virtual MEC system for the 3GPP MEC in order to be transparent.

\iffalse
In order to provide application mobility, there is a need of transferring the session state information from the 3GPP MEC to the fog server. Therefore, in PS3A, proxy acts as a virtual UIE so that fog can request user's session state from the vUIE. The proxy then collects the session state from the 3GPP MEC by acting as the Virtual MEC System. On the other hand, for TSP3A, no additional component in the proxy is required. The virtual components inside the proxy adhere to the specification of the system they are responsible for and ensure a standardized interface to outside world. Here, by virtual we mean that these components are implemented in software modules rather than being implemented on separate hardware. With such implementation, proxy achieves fast, secure, and reliable internal communication.
\fi

\begin{figure}[!t]
\centering
\includegraphics[width=3.6in]{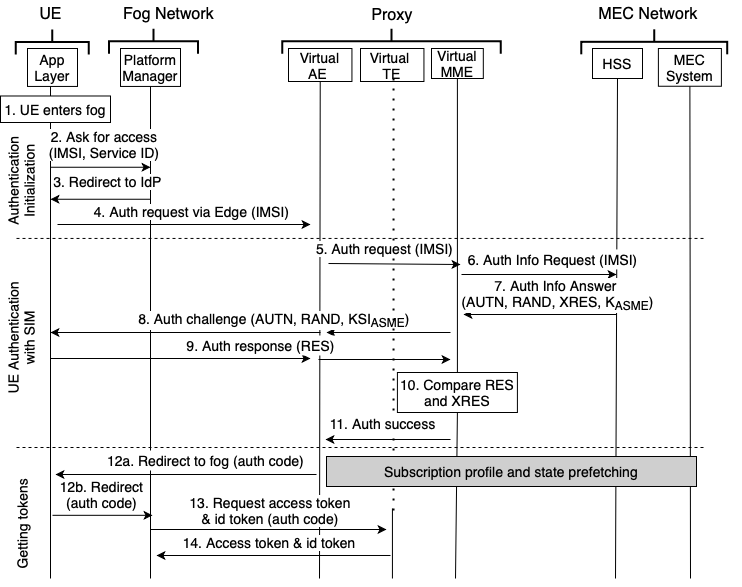}
\caption{Authentication message flow.}
\label{fig_three}
\end{figure}

\subsection{Message Flows}
In order to achieve authentication and application mobility, we identify four stages:

\iffalse
1) \textit{Registration}. A fog and an MEC network make an agreement for federation and register their corresponding components with a proxy agent. 2) \textit{Third-party authentication}: When the user moves from the MEC to the fog network, it is authenticated to fog network with its 3GPP cellular credentials. 3) \textit{Subscription profile collection}: Fog network collects the subscription profile of the user from the MEC and verifies the service access. 4) \textit{Application state transfer}: Application state is transferred from MEC to the fog network and the user resumes using service from the fog network.

After describing these four steps, we discuss how proxy can fetch information early from source MEC to reduce service resumption delay. 
\fi

\subsubsection{Registration} 
The fog and MEC network connect with a proxy agent in this stage. A Diameter connection is set up between the vMME in proxy and HSS in 3GPP cellular network, which is secured via TLS. As per OIDC standards \cite{sakimura2014openid}, the fog platform manager (FPM) in the fog network registers with the vIdP component in the proxy as the Relying Party (RP) in OIDC terms, and receives the client ID.

\subsubsection{Third-Party Authentication}
PS3A and TSP3A authenticate a 3GPP subscriber with the fog network via its 3GPP cellular network credentials in following 3 stages, \textit{authentication initialization, UE Authentication with SIM Credentials, and Obtaining Tokens}, as shown in Fig. \ref{fig_three}. The FPM identifies the user and redirects it to vAE which has it authenticated via vMME and HSS in the 3GPP network by using 3GPP EPS-AKA protocol. After successful authentication, FPM receives an access token and an ID token from vTE for the authenticated user.

\iffalse

The user needs to authenticate with OIDC in fog network via using its 3GPP cellular network credentials. Therefore, we combine the OIDC in fog network and EPS-AKA in the cellular network. The message flow, as shown in Fig. \ref{fig_three}, consists of 3 stages namely: \textit{authentication initialization, UE Authentication with SIM Credentials, and Obtaining Tokens}. PS3A and TSP3A authenticate a 3GPP subscriber with the fog network 

\textit{Authentication Initialization:} When the users enters a fog network, it requests access to a particular service by third-party authentication with cellular credentials. Fog Platform Manager (FPM) identifies the user as the subscriber of the federated 3GPP MEC network and redirects the user to the virtual AE. The UE presents its IMSI to Virtual AE. 

\textit{UE Authentication with SIM Credentials:} 
The user is then authenticated via its SIM credentials. The vIdP sends IMSI internally to the vMME which authenticates the user by using EPS-AKA. The messages between vMME and the user are exchanged via vAE. If the user authentication is successful, vMME informs the vAE about successful authentication.

\textit{Obtaining Tokens:} 
When the vAE confirms successful authentication of the user, it redirects the user to FPM with the authorization code. In the meantime, a new record for the the user containing the IMSI is created in vIdP in proxy. After receiving authorization code, FPM requests access token and ID token from vTE which generates an access token and an ID token and returns these to FPM.  

\fi

\begin{figure}[!t]
\centering
\includegraphics[width=2.9in]{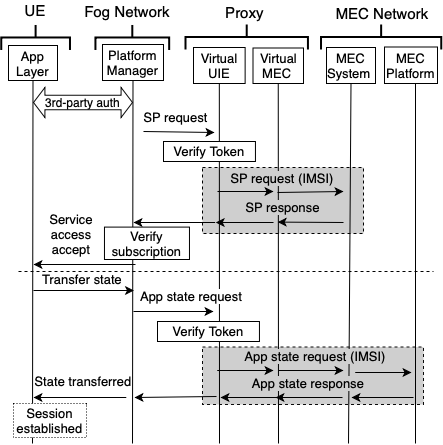}
\caption{Subscription profile collection and state transfer message flow.}
\label{fig_four}
\end{figure}

\subsubsection{Subscription Profile Collection}
After authentication, FPM needs to needs to obtain the user’s subscription profile, stored in the MEC system level \cite{li2020transparent}, to perform authorization, accounting, and ensure QoS. FPM fetches this information from the MEC network and verifies subscription before initializing service application instance for the user as shown in Fig. \ref{fig_four}. Then, FPM sends a subscription profile request to vUIE with the access token which then verifies the token and looks up IMSI for this token. Proxy, while acting as the virtual MEC system, sends this IMSI to the MEC system and collects subscription profile and returns this information to the FPM which verifies the subscription and informs the user whether service access is accepted.

\subsubsection{Application State Transfer}
We use two different methods to transfer the application state from the MEC to the fog platform.

\textit{PS3A.}  
In PS3A, the application state is transferred via the proxy which acts as vUIE to the fog network. The user requests the FPM to transfer the state from the MEC which, in turn, requests vUIE for the application state as a claim with the access token obtained in authentication step. The proxy acts as Virtual MEC system and sends the application state request, with the saved IMSI, to the MEC network which locates the user’s MEC platform via its IMSI and forwards the application state request to that MEC platform. The MEC platform returns the application state to MEC system level, which then returns it to the proxy. The proxy then provides the state to the FPM which initializes the application with this state and initiates a session with the user.

\textit{TSP3A.}
In TSP3A, the UE receives a state token every time it updates its application state while accessing MEC applications. The state token contains the state information and the validity of the token. After disconnecting with the MEC network, the user provides the fog network with the state token after authentication with the fog network. The fog checks the validity of the token and updates the application state, based on the state information within the state token.

\textit{Comparison between PS3A and TSP3A.}
In PS3A, a state is transferred through proxy which adds network delay. TSP3A transfers the state via a token and incurs less delay for the application state transfer. The PS3A transfers the state through backhaul, with the necessary security measures, and TSP3A ensures encryption and integrity protection for the token for secure transfer. In PS3A, fog network always receives the latest state from the MEC while in TSP3A, most up to date state may not be sent and some state information may become lost. Furthermore, a periodic state update in TSP3A adds extra overhead to system. TSP3A is thus suitable for applications that need low latency.

\iffalse
\subsubsection{Prefetching}
After successful authentication, in both PS3A and TSP3A, proxy needs to collect subscription profile from source MEC. Moreover, proxy needs to transfer state from source MEC in PS3A. Proxy can fetch these information early even before the information is requested. These information can be stored in proxy in temporarily and returned to fog network when requested. These two prefetching steps are shown inside rectangles in figure 4. To avoid unnecessary caching, proxy performs prefetching only after third-party authentication is confirmed in Virtual AE as shown in figure 3.

\fi
\section{Implementation}

\begin{figure*}[!t]
\centering
\includegraphics[width=7.2in, height=1.6in]{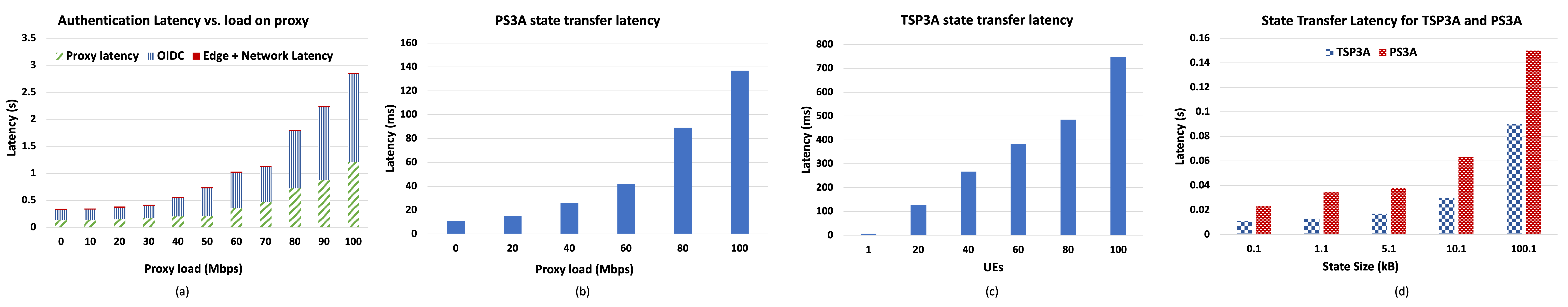}
\caption{(a) Authentication latency, (b) State transfer latency (PS3A), (c) State transfer latency (TSP3A), (d) State transfer latency vs. state size.}
\label{fig_resone}
\end{figure*}

\subsection{Prototype Architecture}
We set up a 3GPP cellular network using open air interface (OAI) and deployed the MEC platform inside a 3GPP cellular network. The MEC application server was implemented using Node.js which simply acted as a backend server for the proxy. We used python and django OIDC for deploying the fog network as an OIDC client. The user equipment (UE) was implemented using python and C. For PS3A, we set up an additional application server in the MEC platform and a state manager, using Node.js, in the fog for handling application state. We also deployed a proxy between the MEC and fog network. The vMME of the proxy was implemented using python and for the vIdP we used sqlite.

\subsection{Testbed}
Our testbed was set up on 2 PCs, both having different specifications and hardware. The user, fog (OIDC client), and the proxy were setup on PC-1 which had a MAC operating system using an i5 processor with 8 GB RAM. The 3GPP cellular network (OAI), together with MEC platform, were deployed on PC-2 which had an Ubuntu operating system running on an i7 processor with 16 GB RAM. The reason for deploying the MEC network separately was to make the MEC network more realistic. Both  PCs were connected to the same LAN through a router/switch.

\section{Results and Evaluation}
For evaluation of our proposed methods, we measured third party authentication latency and state transfer latency for two methods. We also compared the service interruption time taken by our proposed solutions against the authentication and state transfer time taken in the absence of our proposed solution, i.e. via cloud.

\subsection{Authentication Latency}
First, we measured latency due to third-party authentication by applying different loads on the proxy as shown in Fig. 5a. We divided the authentication latency into three parts: 1) proxy latency, 2) OIDC latency, and 3) MEC and network communication latency. The authentication latency was 0.345s without any load on the proxy. We created a network load on proxy by opening hundreds of sockets in the proxy and sending the network traffic to those sockets (base load). The authentication latency significantly increased when the network load was increased on the proxy. Authentication latency increased to 2.858s, about 8 times the delay without any load, with 100 MB/s load. The OIDC latency increased with the proxy load as the proxy acted as a vIdP for the OIDC client i.e the fog and took increasingly more time to serve the OIDC client and thus the increase in OIDC latency. The proxy latency also increased but the increase in OIDC latency was greater than the proxy latency as the proxy prioritizes its own workload in preference to serving the OIDC client.

\iffalse
\begin{figure*}[!t]
\centering
\includegraphics[width=5in]{combined-1.png}
\caption{Authentication latency.}
\label{fig_resone}
\end{figure*}
\fi

\iffalse
\begin{figure}[!t]
\centering
\includegraphics[width=3in, height=2.3in]{res1.png}
\caption{Authentication latency.}
\label{fig_resone}
\end{figure}

\begin{figure}[!t]
\centering
\includegraphics[width=3in, height=2.3in]{res2.png}
\caption{State transfer latency (PS3A).}
\label{fig_restwo}
\end{figure}
\fi

\subsection{State Transfer Latency}
We calculated the state transfer latency for PS3A while increasing the network load on the proxy as shown in Fig. 5b. We used a 100B state, which the proxy was able to pre-fetch from the MEC before it was requested by the fog platform. Therefore, the PS3A state transfer latency was only  as a result of the data transmission from the proxy to the fog platform. Fig. 5a shows that as the network load increased on the proxy, PS3A state transfer latency also increases and ranged between 10.6\textendash137 ms for 0\textendash100 Mbps load. We also analyzed the state transfer latency for TSP3A, as shown in Fig. 5c. In TSP3A, the proxy is not involved in the state transfer and therefore we increased the number of UEs that sent simultaneous state update requests to the fog. TSP3A state transfer latency increased as the number of UEs connected to the fog increased and ranged between 6.5\textendash746.5 ms for 1\textendash100 UEs. The increase in state transfer latency is linear and when the number of UEs reached 100, the state transfer latency increased by a larger amount, most probably as a consequence of the network traffic collision.

\iffalse
\begin{figure}[!t]
\centering
\includegraphics[width=3in, height=2.3in]{res3.png}
\caption{State transfer latency (TSP3A).}
\label{fig_resthree}
\end{figure}

\begin{figure}[!t]
\centering
\includegraphics[width=3in, height=2.3in]{statetransfer.png}
\caption{State transfer latency vs. state size.}
\label{fig_resfour}
\end{figure}
\fi

\subsection{State Transfer Comparison \& Service Interruption Latency}
In order to compare PS3A and TSP3A, we used different state sizes to see how these methods behaved for different state sizes. It can be seen from Fig. 5d that TSP3A state transfer latency ranged between 11\textendash90 ms and PS3A state transfer latency ranged between 23\textendash150 ms for the state size of 0.1\textendash100.1 kB. It should also be noted that, for different state sizes, TSP3A took 40\textendash52\% less time than PS3A because PS3A retrieved the state from an entity (MEC) located further away, whereas TS3A retrieves the state via the UE which took less time. Furthermore, TSP3A state transfer took 94\textendash98\% less time compared to state transfer from the cloud.

We also compared the service interruption latency of PS3A and TSP3A with the state transfer via the cloud. The state size was 100 KB in all three methods. Total service interruption time in TSP3A, PS3A, and cloud-based approach was 0.435s, 0.495s and 2.817s, respectively. TSP3A took the least amount of time, 12.1\% and 84.6\% less than the PS3A and the cloud. PS3A and TSP3A took 82.4\% and 84.6\% less time compared to the cloud, respectively.

\section{Conclusions and Future Work}
A federation among 3GPP MEC and Fog is beneficial for subscribers and providers but it gives rise to third-party authentication and application mobility issues. In this  paper, we proposed PS3A and TSP3A methods that use a proxy for transferring the authentication information of subscribers from 3GPP MEC to the fog, and use proxy and tokens respectively, for the application state transfer. We implemented the proxy on a testbed and the results show that PS3A and TSP3A provide authentication within 0.345\textendash2.858s for 0\textendash100 Mbps proxy load. The results further show that TSP3A provides application mobility while taking 40\textendash52\% less time than PS3A via using state token. TSP3A and PS3A also reduce the service interruption latency by 82.4\% and 84.6\%, compared to  cloud-based service via tokens and prefetching.

This paper addressed the third-party authentication and application mobility problem between the 3GPP MEC and fog for 3GPP subscribers and by proposing a federation among them. In future we will extend this work to include other fog protocols such as 802.1x and PANA. This work can also be extended by considering  IoT device authentication protocols. A federation among different service providers such as cloud, edge, and fog can also be formed in order to create vertical and hybrid federations. This work can also be extended by considering other federation issues, such as resource sharing, accounting, load balancing, and traffic offloading.

\iffalse
\section*{References}

Please number citations consecutively within brackets \cite{b1}. The 
sentence punctuation follows the bracket \cite{b2}. Refer simply to the reference 
number, as in \cite{b3}---do not use ``Ref. \cite{b3}'' or ``reference \cite{b3}'' except at 
the beginning of a sentence: ``Reference \cite{b3} was the first $\ldots$''

Number footnotes separately in superscripts. Place the actual footnote at 
the bottom of the column in which it was cited. Do not put footnotes in the 
abstract or reference list. Use letters for table footnotes.

Unless there are six authors or more give all authors' names; do not use 
``et al.''. Papers that have not been published, even if they have been 
submitted for publication, should be cited as ``unpublished'' \cite{b4}. Papers 
that have been accepted for publication should be cited as ``in press'' \cite{b5}. 
Capitalize only the first word in a paper title, except for proper nouns and 
element symbols.

For papers published in translation journals, please give the English 
citation first, followed by the original foreign-language citation \cite{b6}.
\fi

\vspace{12pt}

\end{document}